\documentclass[a4paper,11pt]{article}
\pdfoutput=1 

\usepackage{jcappub} 

\usepackage{graphicx}
\usepackage[mathlines]{lineno}
\usepackage[english]{babel}
\usepackage{color}
\usepackage{ulem}
\usepackage{enumitem}

\definecolor{amethyst}{rgb}{0.6, 0.4, 0.8}

\def\lapproxeq{\lower .7ex\hbox{$\;\stackrel{\textstyle
<}{\sim}\;$}}
\def\gapproxeq{\lower .7ex\hbox{$\;\stackrel{\textstyle
>}{\sim}\;$}}

\makeatletter
\def\ps@pprintTitle{Published in JCAP as DOI: 10.1088/1475-7516/2019/10/022
 \let\@oddhead\@empty
 \let\@evenhead\@empty
 \def\@oddfoot{}%
 \let\@evenfoot\@oddfoot}
\makeatother

\title{Probing the origin of ultra-high-energy cosmic rays with neutrinos in the EeV energy range using the Pierre Auger Observatory}

\author{The Pierre Auger Collaboration}

\emailAdd{auger\_spokespersons@fnal.gov}

\abstract {Neutrinos with energies above $10^{17}$ eV  are detectable with the Surface Detector Array of the Pierre Auger Observatory. The identification is efficiently performed for neutrinos of all flavors interacting in the atmosphere at large zenith angles, as well as for Earth-skimming $\tau$ neutrinos with nearly tangential trajectories relative to the Earth. No neutrino candidates were found in $\sim\,14.7$ years of data taken up to 31 August 2018. This leads to restrictive upper bounds on their flux. The $90\%$ C.L. single-flavor limit to the diffuse flux of ultra-high-energy neutrinos with an $E_\nu^{-2}$ spectrum in the energy range $1.0 \times 10^{17}~{\rm eV} - 2.5 \times 10^{19}~{\rm eV}$ is $E^2 {\rm d}N_\nu/{\rm d}E_\nu < 4.4 \times 10^{-9}~{\rm GeV~cm^{-2}~s^{-1}~sr^{-1}}$, placing strong constraints on several models of neutrino production at EeV energies and on the properties of the sources of ultra-high-energy cosmic rays.} 


\begin{document}

\maketitle
\flushbottom

\def\linenumberfont{\normalfont\tiny\itshape\sffamily}

\section{Introduction}

The origin, nature and production mechanisms of ultra-high-energy cosmic rays (UHECRs) with energies above $10^{18}$ eV are some of the long-standing questions in astroparticle physics ~\cite{Nagano-Watson_2000, Dawson_2017, Mollerach-Roulet_2018}. The discovery of a large-scale dipolar anisotropy in the arrival directions of UHECRs at energies $E>8 \times 10^{18}$ eV indicates that above this threshold they have an extragalactic origin \cite{Dipole_Science_2017}. The amplitude of the dipole increases steadily with energy above $4 \times 10^{18}$ eV \cite{Dipole_ApJ_2018}.
Indications of anisotropy of UHECRs on intermediate angular scales were also reported recently \cite{Auger_SBG_2018}. The pattern of UHECR arrival directions for energies above $4\times 10^{19}$ eV is matched by a model in which about $10\,\%$ of them arrive from directions that are clustered around bright, nearby galaxies. An isotropic arrival direction distribution of UHECRs was disfavored at 4.0 and 2.7 $\sigma$ by comparing the pattern of arrival directions with flux models based on the local distribution of starburst galaxies and $\gamma$-AGN, respectively. 

Measurements of the UHECR spectrum have revealed a strong suppression of the flux at energies above $\sim\,4\times 10^{19}$ eV with respect to that extrapolated from lower energies \cite{HiRes_spectrum, Auger_spectrum}. The position of the break is compatible with that expected from the Greisen-Zatsepin-Kuzmin (GZK) effect \cite{GZK}, namely the attenuation of the flux of UHECRs due to interactions with the Cosmic-Microwave Background (CMB)  radiation. While the position of the break is compatible with both proton and heavier primaries, data collected with the fluorescence technique at the Pierre Auger Observatory indicate that the composition is getting heavier than protons as the energy increases beyond 2 EeV~\cite{Auger_Xmax}. Models in which medium to heavy nuclei are photodisintegrated while traveling through the Universe \cite{Taylor_2010} cannot be excluded. It is also possible that the suppression is due to a combination of the maximum energy to which sources can accelerate particles and the interactions of the primaries with the background fields~\cite{Allard_2012}. The determination of the composition of the UHECRs \cite{Auger_Xmax, TA_Xmax} is key to distinguishing between these scenarios since they predict different fractions of primaries heavier than protons as energy increases \cite{Allard_2012}. 

All models of UHECR production predict UHE neutrinos as a result of the decay of charged pions generated in interactions of UHECRs within the sources themselves (``astrophysical'' neutrinos), and/or in their propagation through background radiation fields (``cosmogenic'' neutrinos)~\cite{BZ_cosmogenic_1969}. However, the predicted fluxes have large uncertainties, depending strongly on the spatial distribution and redshift ($z$) evolution of the unknown UHECR sources, on the transport model of UHECR assumed, as well as on the spectral features of UHECRs at production~\cite{Kotera_GZK_2010,Heinze19,Alves-Batista19}. Moreover, if UHECRs are heavy nuclei, the UHE neutrino yield is expected to be strongly suppressed~\cite{Hooper_GZK_2005,Ave_GZK_2005}. In this respect the (lack of) observation of UHE neutrinos would provide constraints on the dominant scenario of UHECR production \cite{IceCube_PRL2016} as well as on the evolution with $z$ of their sources, both of  which can help in their identification \cite{Seckel_Stanev_2005,Kotera_GZK_2010}. The fact that neutrinos travel unaffected by magnetic fields and unattenuated implies that EeV neutrinos may be the only direct probe of the sources of UHECRs at distances farther than $\sim\,100$ Mpc.

Astrophysical neutrinos have been observed by the IceCube experiment. Nearly a hundred neutrino events in the energy range $\sim\,100$ TeV and a few PeV have been detected representing a $5.6~\sigma$ excess with respect to atmospheric backgrounds \cite{IceCube_ApJ_2016}. These include four neutrinos of energies above $10^{15}$ eV \cite{IceCube_PRL2014,IceCube_ApJ_2016}. Most remarkable is the recent detection of a $\sim\,3\times10^{14}$ eV neutrino event (IceCube-170922A) in spatial coincidence with the blazar TXS 0506+056 \cite{IceCube_TXS} and in temporal coincidence with a flare of gamma rays detected by the Fermi satellite from the same source. Also, an excess of neutrino events from the position of the blazar was found in the TeV energy range prior to the IceCube event~\cite{IceCube_TXS_prior}. Unfortunately, and despite the variety of techniques being used \cite{Veronique_2011, Alvarez-Muniz_ICRC17}, neutrinos have escaped detection by existing experiments in the EeV energy range \cite{Auger_nus_PRD2015,ANITA_2018,IceCube_PRD2018,ANITA_2019}, i.e.\ about three orders of magnitude above the most energetic neutrinos detected in IceCube. 

In this work, we report on the search for UHE neutrinos in data taken with the Surface Detector Array (SD) of the Pierre Auger Observatory \cite{Auger_observatory_NIMA2015}. A blind scan of data from 1 January 2004 up to 31 August 2018 has yielded no neutrino candidates. This corresponds to $\sim\,9.7$ equivalent years of operation of a complete SD, or $\sim\,14.7$ years of lifetime (because the array was not fully deployed until 2008) representing an increase of $5.2$ years of lifetime, operated with a complete SD, over previous searches \cite{Auger_nus_PRD2015}.
The non-observation of neutrino candidates allows us to place stringent constraints on the diffuse flux of UHE neutrinos with relevant implications for the origin of the UHECRs. In this paper, we report upper limits to the diffuse flux of UHE neutrinos. Corresponding limits to the neutrino flux from point-like sources as a function of declination are the subject of a separate paper \cite{Auger_PS_inprep}.

\section{Searching for UHE neutrinos in Auger data}
\label{sec:selection}

The Pierre Auger Observatory~\cite{Auger_observatory_NIMA2015} is located in the province of Mendoza, Argentina, at a mean altitude of 1400~m above sea level ($\sim\,880~{\rm g~cm^{-2}}$ of vertical atmospheric column density). It was primarily designed to measure extensive air showers (EAS) induced by UHECRs, and has been running and taking data since its construction started in 2004. For that purpose a surface detector (SD)~\cite{Auger_SD_2008, Auger_observatory_NIMA2015} samples the front of shower particles at the ground level with an array of water-Cherenkov detectors (``stations''). Each SD station contains 12 tonnes of water viewed by three nine-inch photomultiplier tubes. The signals produced by the passage of shower particles through the SD stations are recorded as time traces in 25 ns intervals. 1660 SD stations have been deployed over an area of $\sim\,3000~{\rm km^2}$ arranged in a hexagonal pattern with 1.5 km spacing. 

Although the primary goal of the SD of Auger is to detect UHECRs, it can also identify ultra-high-energy (UHE) neutrinos. Neutrinos of all flavors can interact in the atmosphere through charged (CC) or neutral current (NC) interactions and induce a ``downward-going'' (DG) shower that can be detected \cite{Capelle_1998}. The probability of interacting per unit column density traversed is essentially  independent of the atmospheric depth. In addition, $\tau$ neutrinos ($\nu_\tau$) can undergo CC interactions and produce a $\tau$ lepton in the earth crust. The $\tau$ lepton leaves the earth and decays in the atmosphere, inducing an ``Earth-skimming'' (ES) upward-going shower \cite{ES_tau_neutrinos}. Tau neutrinos are not expected to be copiously produced at the astrophysical sources, but as a result of neutrino oscillations over cosmological distances, approximately equal fluxes for each neutrino flavour should reach the earth \cite{Learned_nutau_1995,Athar_nu_oscillations_2000, Anchordoqui_nu_review_2014}. 

Neutrino-induced showers must be identified in Auger data in the background of showers initiated by UHECRs (protons and/or nuclei). The identification is based on a simple idea: highly-inclined (zenith angle $\theta > 60^\circ$), ES and DG neutrino-induced showers initiated deep in the atmosphere near the ground level have a significant electromagnetic component when they reach the SD array, producing signals that are spread over time in several of the triggered SD stations. In contrast, inclined showers initiated by standard UHECRs are dominated by muons at the ground level, inducing signals in the SD stations that have characteristic large peaks associated with individual muons which are spread over smaller time intervals. Thanks to the fast sampling (25 ns) of the SD digital electronics, several observables that are sensitive to the time structure of the signal can be used to discriminate between these two types of showers. 
Auger data are searched for UHE neutrino candidates in both the DG and ES analyses, in the zenith angle ranges $60^\circ<\theta<90^\circ$ and $90^\circ<\theta<95^\circ$ respectively.

\subsection{General search strategy}

The search strategy consists in selecting showers that arrive at the SD array in the inclined directions and identifying those that exhibit a broad time structure in the signals induced in the SD stations. Such signals are indicative of an early stage of development of the shower, a signature of the  shower developing close to the ground. To define the selection algorithms and optimize the numerical values of the cuts needed to identify neutrino-induced showers we follow a blind analysis procedure. A fraction of $\sim\,15\,\%$ of the whole data sample (from 1 January 2004 up to 31 August 2018), along with Monte Carlo simulations of UHE neutrinos, is dedicated to define the selection algorithms, the most efficient observables for neutrino identification, and the value of the cuts. This ``training" data set is assumed to be constituted of background UHECR-induced showers. The remaining fraction of data (``search data") is ``unblinded" to search for neutrino candidates but only after the selection procedure is established. 

The selection of inclined showers is tailored to the different zenith angle ranges where the search is performed, namely DG and ES. Since standard angular reconstruction techniques have larger uncertainties for nearly horizontal events~\cite{Auger_HAS_recons_2014,Tifenberg_PhD_2011}, another strategy to select inclined events has been followed.
For purely geometrical reasons, in inclined events the pattern of the
triggered SD stations typically exhibits an elliptical shape on the ground with the major axis of the ellipse along the azimuthal arrival direction. These patterns can be characterized by a length $L$ (major axis) and a width $W$ (minor axis). Inclined events typically exhibit large values of $L/W$, and an appropriate cut in $L/W$ is therefore an efficient selector of inclined events \cite{Auger_ES_PRL2008, Auger_ES_PRD2009, Auger_DGH_PRD2011}. Another indication of the arrival direction of the event is given by the average (apparent) speed $\langle V \rangle$ of the trigger time from one station to another. This is calculated from the projected distance between pairs of stations along the major axis of the ellipse and the trigger times, and it is averaged over all pairs of stations in the event. In vertical showers $\langle V \rangle$ exceeds the speed of light since all triggers occur at roughly the same time, while in very inclined events $\langle V \rangle$ is close to the speed of light. In addition, the Root-Mean-Square (RMS($V$)) of the apparent speed (as obtained from the values of $V$ using different pairs of stations) is typically below $\sim\,25\%$ of $\langle V \rangle$ \cite{Auger_ES_PRL2008, Auger_ES_PRD2009, Auger_DGH_PRD2011}.

\begin{figure}[ht]
\centering
\includegraphics[width=0.9\textwidth]{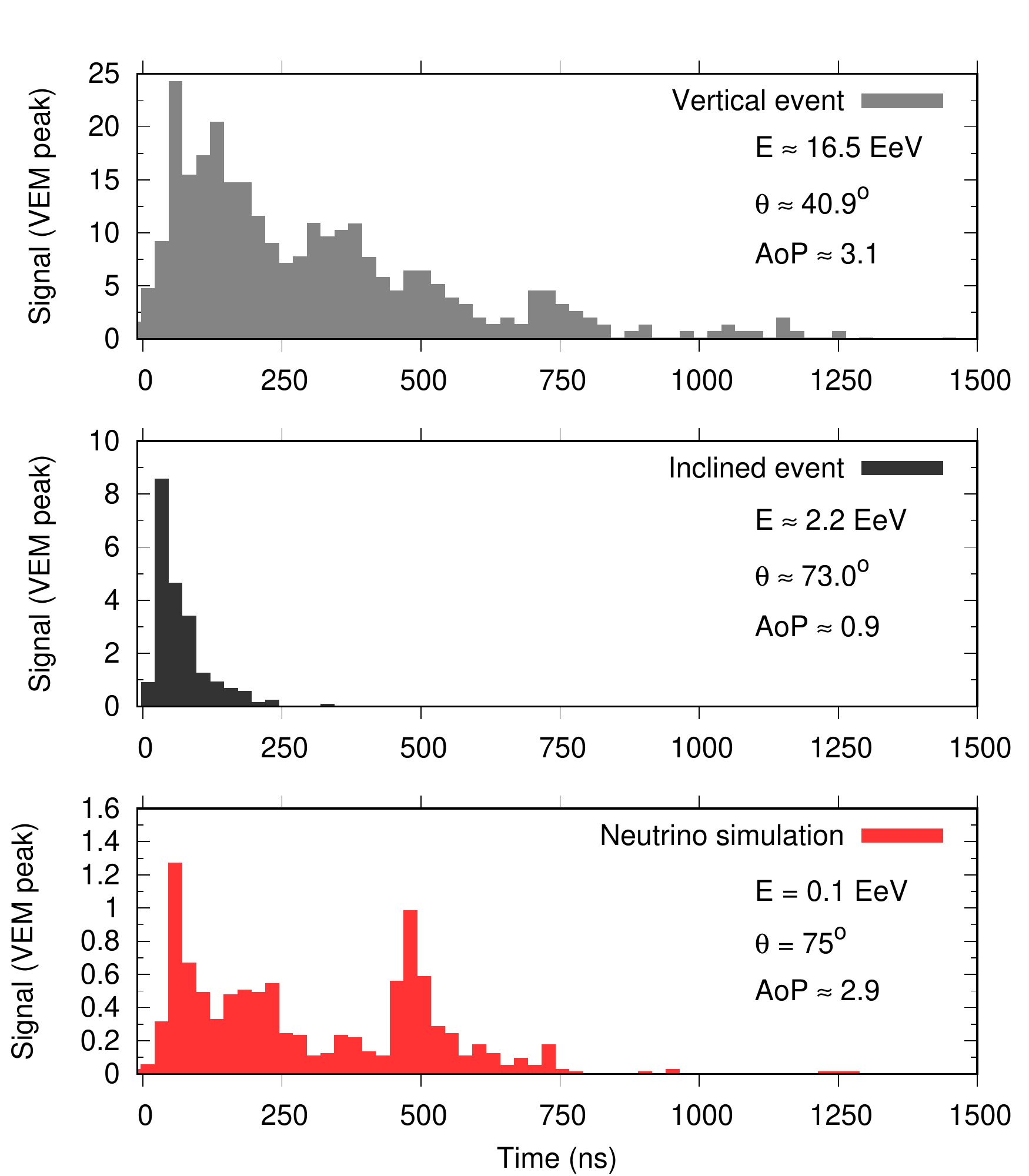}
\caption{
FADC traces of stations at a distance of approximately 1 km to the shower core. From top to bottom panel the station belongs to a vertical event, to an inclined event, and to a neutrino-simulated event. The reconstructed energy ($E$) and zenith angle ($\theta$) for the events, as well as the simulated $E$ and $\theta$ of the neutrino-induced shower, are indicated in each panel. The value of the Area-over-Peak (AoP) of each trace is also given.
}
\label{fig:traces}
\end{figure}

For the purpose of identifying those inclined events that interact deep in the atmosphere, several observables that contain information on the spread in time in the SD stations can be extracted from the time traces. Among them the Area-over-Peak\footnote{The Area-over-Peak is defined as the ratio of the integral of the time trace to its peak value normalized to the average signal produced by a single vertical muon.} (AoP) has been shown to serve as an efficient observable to discriminate broad from narrow shower fronts \cite{Auger_nus_PRD2015}. Inclined background showers of hadronic origin, in which the muons arrive at the SD stations in a short time interval, exhibit AoP values close to one by definition corresponding to the average AoP of single vertical muons used for calibration. 
In contrast in neutrino-induced showers the values of AoP are typically larger. 
This can be seen in Fig.~\ref{fig:traces} where we show examples of traces of stations belonging to a vertical and an inclined event detected with the SD of Auger, as well as the trace of a station in a neutrino-induced simulated event. 
The optimization of the algorithms based on AoP and related observables to identify deeply-initiated showers is done separately in each angular range as described briefly in the following and in more detail in \cite{Auger_nus_PRD2015}.  

\subsection{Earth-skimming (ES) neutrinos} 

The algorithms to reconstruct the zenith angle of arrival of the primary particles in Auger \cite{Auger_HAS_recons_2014} have been designed for showers that have an impact point within the array. However, a $\nu_\tau$-induced ES shower travels nearly parallel to the earth's surface, in a slightly upward direction, and its core does not intersect the ground. The shower particles deviate laterally from the axis and can reach the ground triggering the SD stations. For this reason, the selection of inclined ES showers is just based on the properties of the signal pattern obtained with the triggered SD stations and the apparent propagation speed of the signal at ground level.  Above the energy threshold of the SD (around $10^{17}$ eV), $\tau$ leptons are efficiently produced only at zenith angles between $\theta=90^\circ$ and $\theta=95^\circ$ \cite{Payet_PRD2008,NuTauSim_2018}. Moreover, from Monte Carlo simulations of $\tau$-lepton-induced showers we have determined that the trigger efficiency of the SD decreases rapidly above $\theta\sim95^\circ$.
For these reasons, we place restrictive cuts to select quasi-horizontal showers with highly elongated signal patterns, namely, $L/W>5$ and  $\langle V \rangle \in [0.29, 0.31]~{\rm m\,ns^{-1}}$ with RMS($V$)$\,<0.08~{\rm m\,ns^{-1}}$ (see also Table~I in \cite{Auger_nus_PRD2015}).
In the ES analysis the search is performed in all events with at least 3 triggered stations $N_{\rm stat}\geq 3$. 

For data prior to 31 May 2010, the neutrino identification variables included the fraction of stations with Time-over-Threshold (ToT) \cite{Auger_trigger_2010} trigger\footnote{This is a type of trigger designed to select spread-in-time sequences of small signals in the time traces \cite{Auger_trigger_2010}.} and having AoP$\,>1.4$~\cite{Auger_ES_PRL2008, Auger_ES_PRD2009}. This fraction is required to be above $60\,\%$ of the triggered stations in the event. For data beyond 1 June 2010, an improved selection is adopted using the average value of AoP ($\langle{\rm AoP}\rangle$) over all the triggered stations in the event as the only observable to discriminate between hadronic showers and ES neutrinos. The value of the cut on $\langle$AoP$\rangle$ is fixed using the tail of the distribution of $\langle$AoP$\rangle$ in real data, which is consistent with an exponential function. This tail is fitted and extrapolated to find the value of $\langle$AoP$\rangle$ corresponding to less than 1 expected background event per 50 yr on the full SD array (see \cite{Auger_nus_PRD2015} for full details). 

\begin{figure}[ht]
\centering
\includegraphics[width=0.9\textwidth]{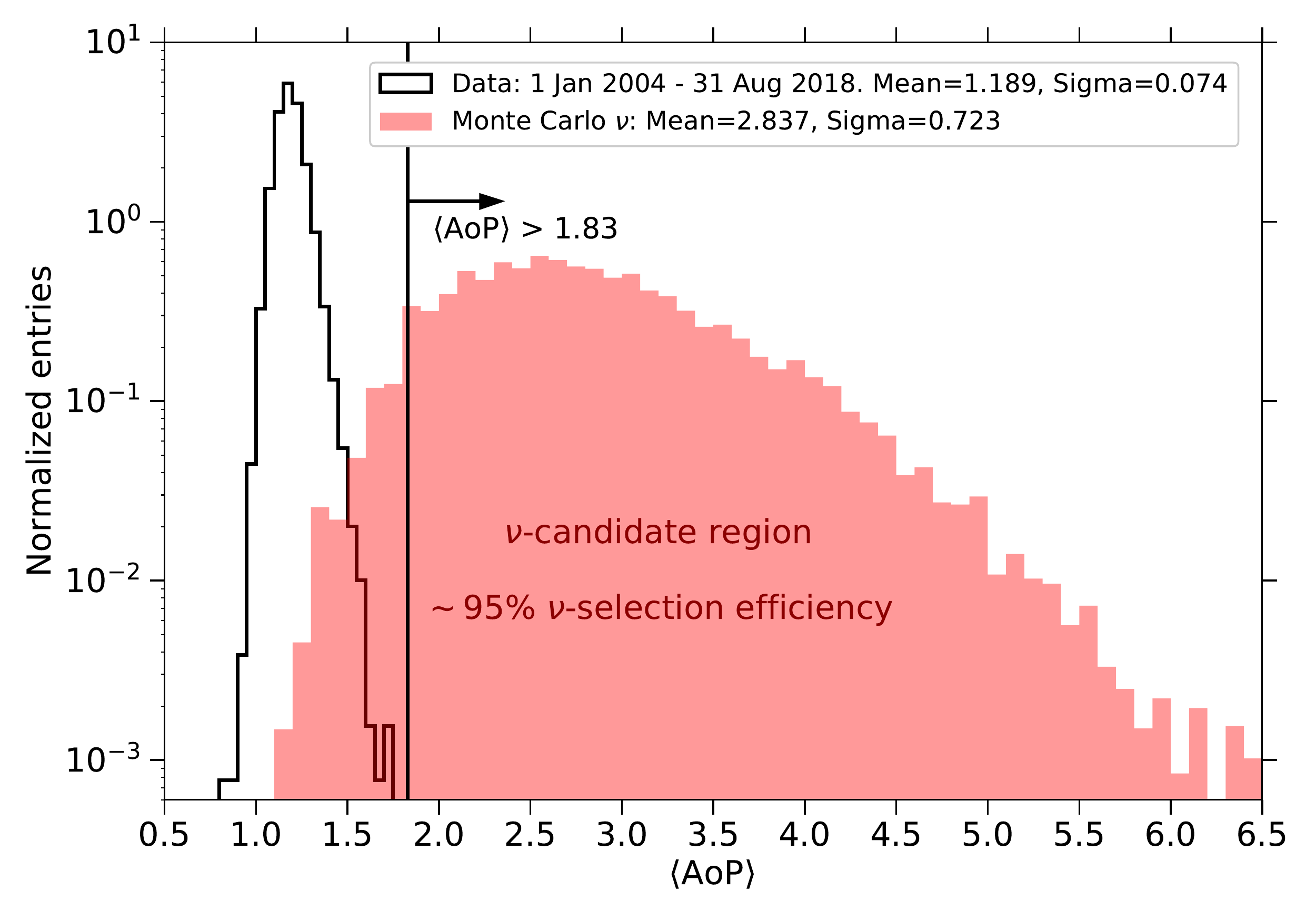}
\caption{
Distribution of $\langle {\rm AoP} \rangle$ after the Earth-skimming inclined selection. Black histogram: full data set up to 31 August 2018 containing 25904 events. Red-shaded histogram: Monte Carlo simulated ES $\nu_\tau$ events. 
}
\label{fig:AoP}
\end{figure}

Applying these criteria, a search for ES neutrino-induced showers is performed in the Observatory data from 1 January 2004, when data taking started, up to 31 August 2018. No neutrino candidates are identified. In Fig.~\ref{fig:AoP} we show the distribution of $\langle$AoP$\rangle$ for the whole data period  compared to that expected in Monte Carlo simulations of $\nu_\tau$-induced ES showers, along with the optimized value of the cut ($\langle$AoP$\rangle=1.83$) above which an event would be regarded a neutrino candidate. After the inclined selection and the neutrino identification criteria, $\sim\,95\%$ of the simulated neutrinos that induce triggers are kept. This proves that the Pierre Auger Observatory is highly efficient as a neutrino detector, with its sensitivity mostly governed by its lifetime and the available target matter for neutrino interactions along the earth's chord. The neutrino search is not limited by the background due to UHECR-induced showers since this can be very efficiently reduced as shown in Fig.~\ref{fig:AoP}.

\subsection{Downward-going (DG) neutrinos} 

For optimization purposes, the DG category of events is further subdivided into two sets for Low (DGL) and High (DGH) zenith angles, between $60^\circ<\theta<75^\circ$ \cite{Auger_nus_PRD2015, Navarro_PhD_2012} and $75^\circ<\theta<90^\circ$ \cite{Auger_DGH_PRD2011} respectively.

Since the core of DG showers always hits the ground, standard angular reconstruction techniques \cite{Auger_HAS_recons_2014} can be used to obtain an estimate of the zenith angle of the shower. However, these techniques have larger uncertainties for nearly horizontal events~ \cite{Auger_HAS_recons_2014,Tifenberg_PhD_2011}. For this reason the primary observables for inclined selection in the DGH case are the ratio $L/W$ of the signal pattern of the shower at ground as well as the apparent average velocity of the signal $\langle V\rangle$, in addition to a simple estimate of the zenith angle $\theta_{\rm rec}$ \cite{Auger_nus_PRD2015,Tifenberg_PhD_2011}. In the case of DGH showers the cuts on the properties of the signal pattern are $L/W>3$, $\langle V \rangle <0.313~{\rm m\,ns^{-1}}$ and RMS($V$)/$\langle V \rangle<0.08$, along with a further requirement on the estimated shower zenith angle $\theta_{\rm rec}>75^\circ$ (see Table I in \cite{Auger_nus_PRD2015}). In contrast, in the DGL case, corresponding to $60^\circ < \theta < 75^\circ$, restrictions on the signal patterns have been found to be less efficient in selecting inclined events than $\theta_{\rm rec}$ \cite{Navarro_PhD_2012}, and only an angular cut $58.5^\circ < \theta_{\rm rec} \leq 76.5^\circ$ is applied, including some allowance to account for the resolution in the angular reconstruction of the simulated neutrino events \cite{Navarro_PhD_2012}. In both the DGH and DGL cases, at least 4 stations $(N_{\rm stat}\geq 4)$ are required in the event. 

For DG showers, we use a multivariate analysis (Fisher method \cite{Fisher}) combining several observables that carry information on the time spread of the signals in the SD stations. These observables are constructed from the AoP values of individual stations. This analysis is optimized with Monte Carlo simulations of UHE neutrinos and the training sample of real data. 
Based on the information obtained from simulations the selection is further improved when dividing the DGH category of events into three sets depending on the number of triggered stations $N_{\rm stat}\leq 6$, $7\leq N_{\rm stat}\leq 11$ and $N_{\rm stat}\geq 12$. Correspondingly for the DGL category the division is in five sets depending on the reconstructed zenith angle $58.5^\circ < \theta_{\rm rec}\leq 61.5^\circ, ~61.5^\circ < \theta_{\rm rec} \leq 64.5^\circ, ~64.5^\circ < \theta_{\rm rec} \leq 67.5^\circ,~67.5^\circ < \theta_{\rm rec} \leq 70.5^\circ$, and $70.5^\circ < \theta_{\rm rec} \leq 76.5^\circ$. A different Fisher discrimination variable is constructed in each subcategory and the cut value independently optimized. 

In the DGH channel, the linear Fisher discriminants are constructed with ten variables that exploit the fact that, due to the large inclination of the shower, the electromagnetic component is less attenuated in the stations that are first hit by a deep inclined shower than in those that are hit last \cite{Auger_DGH_PRD2011,Tifenberg_PhD_2011}. Using Monte Carlo simulations of $\nu-$induced showers with $\theta > 75^\circ$ we have found that in the first few stations hit in the event the AoP values range between 3 and 5, while AoP is typically $\sim\,1$ in stations triggered later. Based on this we have found good discrimination when constructing the Fisher discriminant with the AoP and (AoP)$^2$ of the four stations that trigger first in each event, the product of the four AoPs, and a global parameter of the event that is sensitive to the asymmetry between the average AoP of the early stations and those triggering last in the event (see \cite{Auger_nus_PRD2015} for further details).
 
The selection of neutrino candidates in the zenith angle range DGL category is more challenging because the electromagnetic component of UHECR-induced showers at the ground increases as the zenith angle decreases, imposing a larger background to the searches for neutrinos. Based on the information from Monte Carlo simulations in this angular range, it has been determined that out of all triggered stations of an event, the ones closest to the shower core exhibit the highest discrimination power in terms of AoP. 
In fact, it has been found that the first triggered stations can still contain some electromagnetic component for background events and, for this reason, they are not used for discrimination purposes. The last ones, even if they are triggered only by muons from a background cosmic-ray shower, can exhibit large values of AoP because they are far from the core where muons are known to arrive with a larger spread in time. The variables used in the Fisher discriminant analysis in the DGL channel are the individual AoP of the four (five) stations closest to the core for events with $\theta_{\rm rec}\leq 67.5^\circ$ ($\theta_{\rm rec}>67.5^\circ$) and their product \cite{Auger_nus_PRD2015, Navarro_PhD_2012}. Finally, in the DGL analysis it is also required that at least $75\,\%$ of the triggered stations closest to the core have a ToT local trigger \cite{Auger_nus_PRD2015, Navarro_PhD_2012}.

Once the Fisher discriminant $\cal F$ is defined, an optimized value for the cut is selected to efficiently separate  neutrino candidates from regular hadronic showers for both the DGH and DGL channels. The value is chosen performing an exponential fit to the Fisher distribution of the training data, extrapolating it, and requiring less than 1 event per 50 yr on the full SD array for each multiplicity sub-sample in the DGH channel, and 1 event per 20 yr in the DGL channel \cite{Auger_nus_PRD2015, Navarro_PhD_2012}.

Applying these criteria, a search for DG neutrino-induced showers is performed in the Observatory data since 1 January 2004 up to 31 August 2018. No neutrino candidates are identified in any of the three subcategories of DGH. 

\begin{figure}[ht]
\centering
\includegraphics[width=0.9\textwidth]{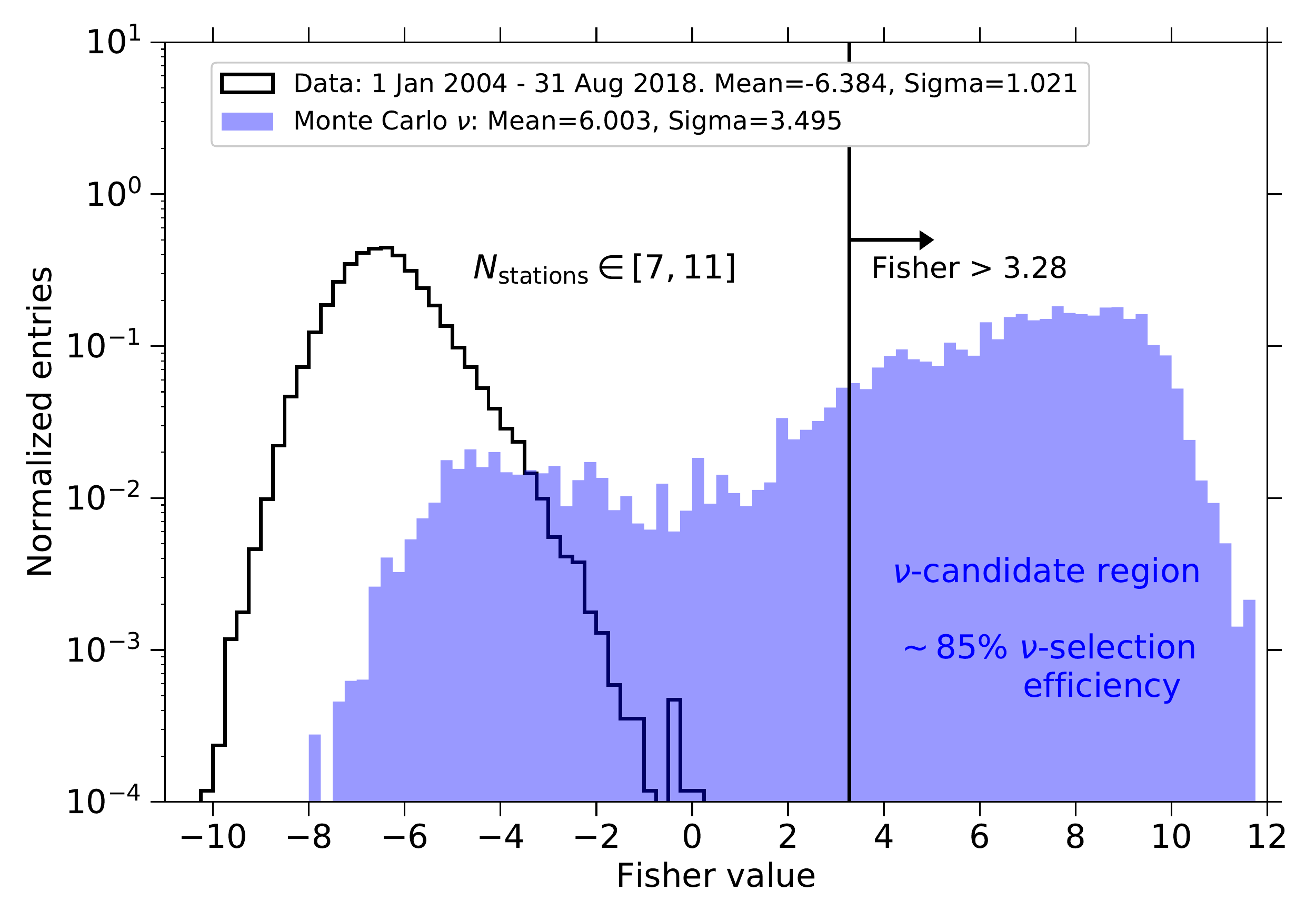}
\vskip -3mm
\caption{
Distribution of the Fisher variable after the downward (DGH) inclined event selection for events with number of triggered stations between 7 and 11. Black histogram: full data set up to 31 August 2018 containing 33885 events. 
Blue-shaded histogram: Monte Carlo simulated $\nu$ DGH events. 
}
\label{fig:Fisher_DGH_med}
\end{figure}

In Fig.~\ref{fig:Fisher_DGH_med}, we show the distribution of the Fisher variable in the DGH subsample with $N_{\rm stat}\in [7,~11]$ for the whole data period compared to that expected in Monte Carlo simulations of $\nu$-induced DGH showers, along with the optimized value of the cut above which an event would be regarded a neutrino candidate. Of all the simulated inclined $\nu$ events that trigger the Observatory, a fraction between $\sim\, 81\,\%$ and $\sim 85\,\%$ (depending on $N_{\rm stat}$) are kept after the cuts on the Fisher variable in the DGH analysis. 

\begin{figure}[ht]
\centering
\includegraphics[width=0.9\textwidth]{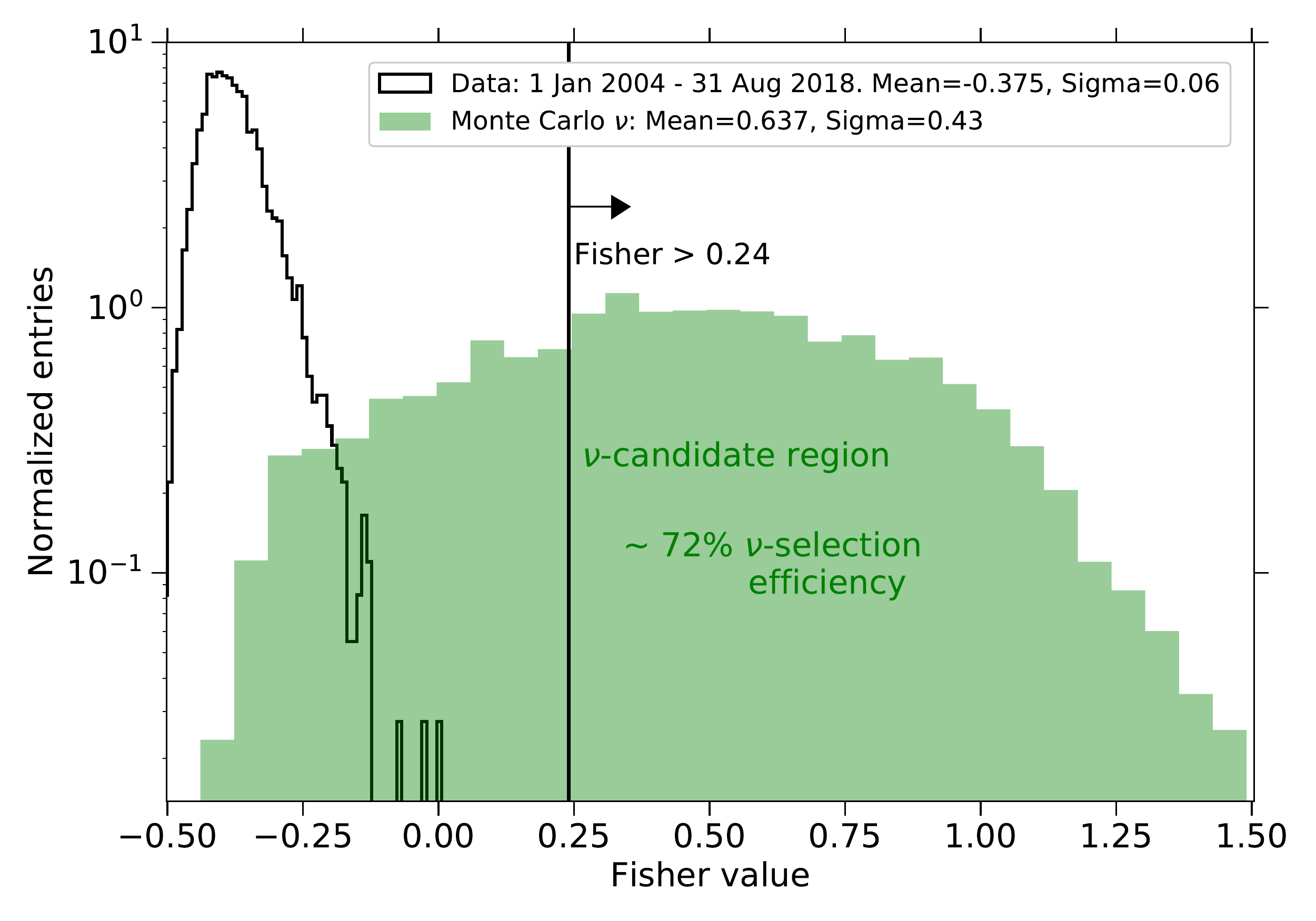}
\vskip -3mm
\caption{
Distribution of the Fisher variable after the downward (DGL) inclined event selection for events with reconstructed zenith angle $64.5^\circ<\theta\leq{}67.5^\circ$. Black histogram: full data set up to 31 August 2018 containing 3948 events.
Green-shaded histogram: Monte Carlo simulated $\nu$ DGL events.
}
\label{fig:Fisher_DGL_3}
\end{figure}

No candidates are found in any of the five DGL subcategories during the same period. As an example, we show in Fig.~\ref{fig:Fisher_DGL_3} the distribution of the Fisher variable in the DGL subset with $\theta_{\rm rec}\in (64.5^\circ,~67.5^\circ]$ and that obtained in Monte Carlo simulations of $\nu$-induced DGL showers in the same zenith angle range. In this case, a fraction between $\sim\,51\,\%$ and $\sim\,77\,\%$ of the simulated inclined $\nu$ events that trigger are kept after applying the cuts on the Fisher variable, with the fraction depending on the $\theta_{\rm rec}$ range considered, the lowest fraction achieved at the smallest values of $\theta_{\rm rec}$. The lower identification efficiency on average in the DGL selection when compared to that in DGH, is due to the more stringent criteria in the angular bin $\theta \in (58.5^\circ, 76.5^\circ]$ needed to reject the larger contamination from cosmic-ray induced showers.

\section{Exposure}
\label{sec:exposure}

The non-observation of neutrino candidates can be converted into an upper limit to the diffuse flux of UHE neutrinos. For this purpose, the exposure of the SD of Auger needs to be calculated for the period of data taking. This is done with Monte Carlo simulations of neutrino-induced showers. The same selection and identification criteria applied to the data were also applied to the results of these simulations. The identification efficiencies for each channel were obtained as the fraction of simulated events that trigger the Observatory and pass the selection procedure and identification cuts~\cite{Auger_DGH_PRD2011}. The results presented here can be obtained after solid angle integration of the directional exposure for point-like sources addressed in~\cite{Auger_PS_inprep}. 

\begin{figure*}[ht]
\centering
\includegraphics[width=0.9\textwidth]{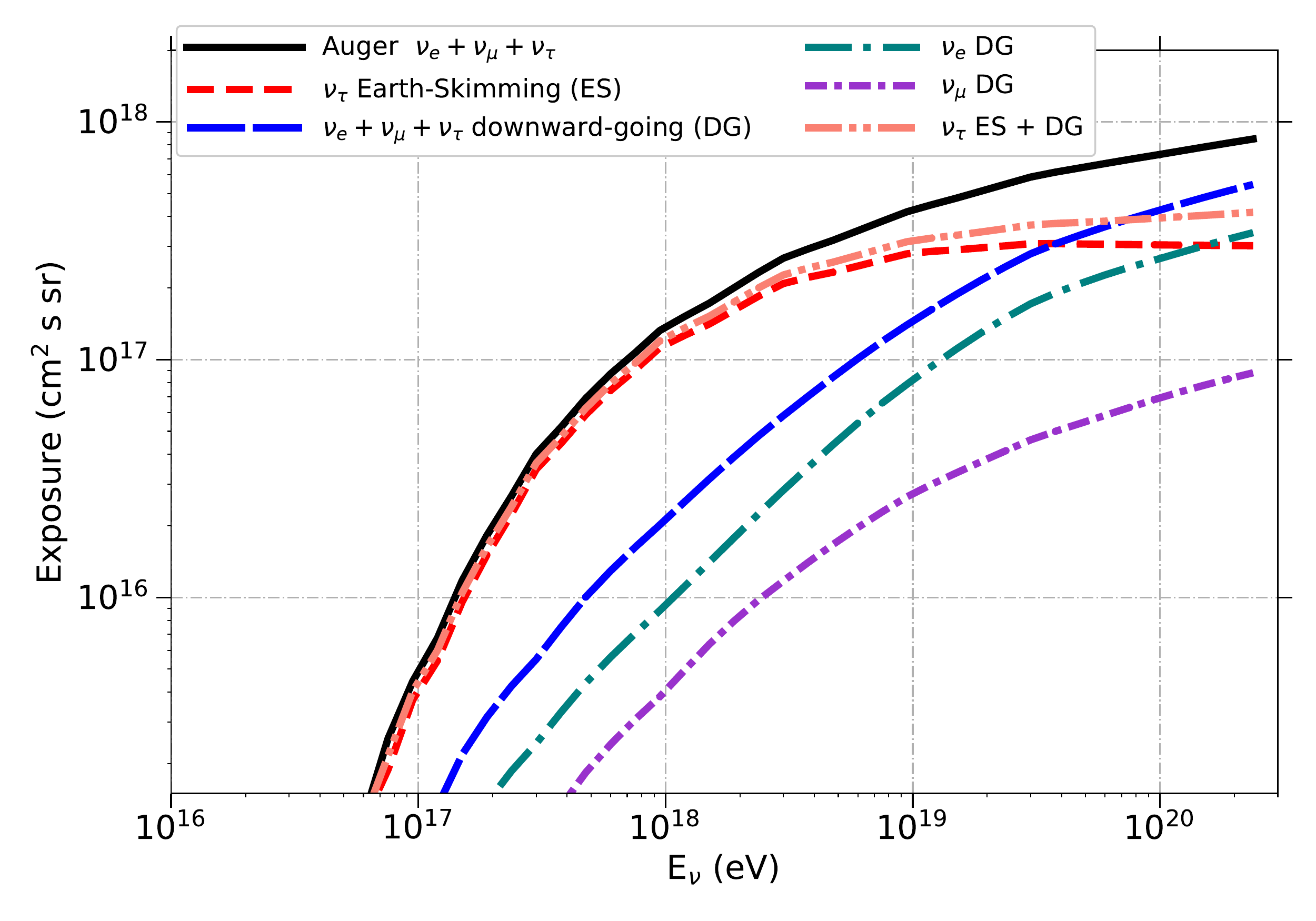}
\caption{
Exposure of the SD of the Pierre Auger Observatory (1 January 2004 - 31 August 2018) to UHE neutrinos as a function of neutrino energy for each neutrino flavor and for the sum of all flavors assuming a flavor mixture of $\nu_e:\nu_{\mu}:\nu_\tau=1:1:1$. Also shown are the exposures to upward-going Earth-skimming $\nu_\tau$ only and to the Downward-Going neutrinos of all flavors including CC and NC interactions. 
}
\label{fig:exposure}
\end{figure*}

For downward-going neutrinos, the detection efficiency, $\varepsilon_{i,c}$, depends on neutrino flavor $i=\nu_e,\nu_\mu,\nu_\tau$, the type of interaction ($c=$CC, NC), neutrino energy $E_\nu$, zenith $\theta$ and azimuth $\varphi$ angles, the point of impact of the shower core on the ground, and the depth in the atmosphere $X$ measured along the shower axis at which the neutrino is forced to interact in the simulations \cite{Auger_nus_PRD2015, Auger_DGH_PRD2011}. The detection efficiency also has some dependence on time because the SD array grew steadily from 2004 up to 2008 when it was completed, and because the fraction of working stations - typically above $95\%$ - is changing continuously with time. For downward-going neutrinos the calculation of the exposure involves folding the detection efficiencies with the area of the SD projected onto the direction perpendicular to the arrival direction of the neutrino, and with the $\nu$ interaction probability for a neutrino energy $E_\nu$ at a depth $X$
which also depends on the type of interaction (CC or NC). Integrating $\varepsilon$ over the parameter space ($\theta,~\varphi,~X$), detector area, and time over the search period, yields the exposure for flavor $i$ and channel $c$~\cite{Auger_nus_PRD2015, Auger_DGH_PRD2011},
\begin{equation}
{\cal E}_{i,c,\rm DG}(E_\nu)= 
~\int_A~\int_\theta~\int_\varphi~\int_X~\int_t~\cos\theta \sin\theta~ 
\varepsilon_{i,c}~\sigma^c_{\nu}~m_{\rm p}^{-1}~{\rm d}A~{\rm d}\theta~{\rm d}\varphi~{\rm d}X~{\rm d}t.
\label{eq:exposure_DG}
\end{equation}
The term $\sigma^c_{\nu}~m_p^{-1}~{\rm d}X$ , with $m_p$ the mass of a proton, and $\sigma^c_\nu$ the neutrino-nucleon cross-section~\cite{Cooper-Sarkar_2011}, represents the probability of neutrino-nucleon interactions along a depth ${\rm d}X$ (in g\,cm$^{-2}$).
An effective exposure can be obtained by summing over channels (CC and NC) and neutrino flavors, weighting each with the relative flavor ratios (for instance weight 1 for each flavor corresponding to a 1:1:1 flavor ratio at earth). 

In the $\nu_\tau$ Earth-skimming channel, the calculation of the exposure is more involved. The efficiency $\varepsilon_{\rm ES}$  depends on the energy of the emerging $\tau$ leptons $E_\tau$, the position of the signal pattern on the ground, on the zenith $\theta$ and azimuth $\varphi$ angles, and on the altitude $h_{\rm dec}$ of the decay point of the $\tau$ above ground \cite{Auger_ES_PRD2009}. $\varepsilon_{\rm ES}$ is averaged over the decay channels of the $\tau$ with their corresponding branching ratios, that determine the amount of energy fed into the shower. The detection efficiency $\varepsilon_{\rm ES}$ is folded with the projected area of the SD, with the differential probability $p_{\rm exit}(E_\nu,\theta,E_\tau)={\rm d}p_{\rm exit}/{\rm d} E_{\tau}$ of a $\tau$ emerging from the earth with energy $E_\tau$ (given a neutrino with energy $E_\nu$ crossing an amount of earth determined by the zenith angle $\theta$), as well as with the differential probability $p_{\rm dec}(E_\tau,\theta,h_{\rm dec}) = {\rm d}p_{\rm dec}/{\rm d}h_{\rm dec}$ that the $\tau$ decays at an altitude $h_{\rm dec}$ \cite{Auger_ES_PRD2009}. An integration over the whole parameter space ($E_\tau,~\theta,~\varphi,~h_{\rm dec}$), area, and time yields the exposure \cite{Auger_nus_PRD2015, Auger_ES_PRD2009}:
\begin{equation}
{\cal E}_{\rm ES}(E_\nu)= 
\int_A~\int_\theta~\int_\varphi~\int_{E_\tau}~\int_{h_{\rm dec}}~\int_t~ \vert \cos\theta\vert~\sin\theta~   
p_{\rm exit}~p_{\rm dec}~\varepsilon_{\rm ES}~
{\rm d}A~{\rm d}\theta~{\rm d}\varphi~{\rm d}E_{\tau}~{\rm d}h_{\rm dec}~{\rm d}t.
\label{eq:exposure_ES}
\end{equation}
The probability $p_{\rm exit}(E_\nu,E_\tau,\theta)$ is calculated using a dedicated Monte Carlo simulation of $\nu_\tau$ propagation inside the earth \cite{Payet_PRD2008} that includes $\nu_\tau$ interaction through CC (NC) channels and sampling of the resulting $\tau$ ($\nu_\tau$) energy, $\tau$ energy loss, and $\tau$ decay inside the earth before emerging with $\nu_\tau$ regeneration following the decay \cite{Payet_PRD2008, NuTauSim_2018}. The probability of $\tau$ decay after traveling a distance $l=h_{\rm dec}/\vert\cos\theta\vert$ can be obtained analytically $p_{\rm dec}(E_\tau,\theta,h_{\rm dec}) = (\lambda_{\rm dec}\vert\cos\theta\vert)^{-1} ~ \exp{(-l/{\lambda_{\rm dec}})}$ where $\lambda_{\rm dec} = \gamma_\tau c\tau_0 = E_\tau c\tau_0/m_\tau$ is the average $\tau$ decay length with $m_\tau\simeq 1.777$ GeV the mass of the $\tau$ lepton, and $c\tau_0\simeq 86.93\times10^{-6}$ m the lifetime of the $\tau$ multiplied by the speed of light $c$.  

The total exposure, ${\cal E}_{\rm tot}$, obtained assuming a flavor mixture of $\nu_e:\nu_{\mu}:\nu_\tau=1:1:1$ is plotted in Fig.~\ref{fig:exposure}. Also shown are the exposures for each neutrino flavor ${\cal E}_i$ with $i=\nu_e,\nu_{\mu},\nu_\tau$, as well as the separate contribution to the exposure due to Earth-skimming $\nu_\tau$ alone and the sum of all the exposures due to the three flavors and two channels (CC \& NC) contributing to the downward-going neutrino category.

When comparing the different neutrino flavors, the $\nu_\tau$ dominates the contribution to the total exposure mainly because of the enhanced sensitivity of the Earth-skimming channel that stems from the denser target provided by the earth's crust for the ES $\nu_\tau$ to interact, and the large range of $\tau$ leptons at EeV energies ($\sim\,10$ km at 1 EeV in rock \cite{Seckel_PRD2001}). The contribution to the $\nu_\tau$-flavor exposure ${\cal E}_{\nu_\tau}$ due to downward-going $\nu_\tau$ is typically small, but increases as the energy rises reaching a fraction of $\sim\,30\,\%$ of ${\cal E}_{\rm ES}$ at $E_\nu=10^{20}$ eV. This can be seen in Fig.~\ref{fig:exposure} by comparing ${\cal E}_{\nu_\tau}$ and ${\cal E}_{\rm ES}$. 
Next in importance is the exposure due to electron neutrinos ${\cal E}_{\nu_e}$ (CC and NC interactions together) dominating over that due to muon neutrinos ${\cal E}_{\nu_\mu}$. This is partly due to the larger fraction of neutrino energy going into the shower in $\nu_e$ CC interactions as compared to $\nu_\mu$ CC interactions, and partly due to the larger electromagnetic content of the shower induced by the electron produced in the $\nu_e$ CC interaction. When comparing the exposure for ES $\nu_\tau$ with the sum of all flavors and channels contributing to the downward-going exposure, it can be seen in Fig.~\ref{fig:exposure} that the latter dominates only above $\sim\,3-4 \times 10^{19}$ eV. Above this energy the exposure to ES $\nu_\tau$ starts to decrease due to the growing decay length of the emerging $\tau$ leptons in air at these energies, favouring decays at high altitude above the detector and reducing the probability that the $\tau$-induced shower triggers the SD array.

\section{Limits to diffuse fluxes}
\label{sec:results}

The total exposure ${\cal E}_{\rm tot}$ folded with a single-flavor flux of UHE neutrinos per unit energy, area $A$, solid angle $\Omega$ and time, $\phi(E_\nu)={\rm d}^6N_\nu/({\rm d}E_\nu~{\rm d}\Omega~{\rm d}A~{\rm d}t)$ and integrated in energy gives the expected number of events for that flux:
\begin{equation}
N_{\rm evt} = ~\int_{E_\nu}~{\cal E}_{\rm tot}(E_\nu)~\phi(E_\nu)~{\rm d}E_\nu.
\label{eq:rate}
\end{equation}
Assuming a differential neutrino flux $\phi = k\cdot E_\nu^{-2}$, an upper limit to the value of $k$ at $90\%$ C.L. is obtained as
\begin{equation}
k_{90} = \frac{2.39}{\int_{E_{\nu}}~E_{\nu}^{-2}~{\cal E}_{\rm tot}(E_\nu)~{\rm d}E_\nu},
\label{eq:k90a}
\end{equation}
where $2.39$ is the Feldman-Cousins factor \cite{Feldman-Cousins} for non-observation of events in the absence of expected background accounting for systematic uncertainties \cite{Conrad,Auger_nus_PRD2015}. The integrated limit represents the value of the normalization of a $E_\nu^{-2}$ differential neutrino flux needed to predict $\sim\,2.39$ expected events. 

Several sources of systematic uncertainty have been considered in the calculation of the exposure and limit. The uncertainty due to simulations includes the effects of using several neutrino interaction generators, shower simulations, hadronic interaction models and thinning level. These would modify the event rate for a $\phi(E_\nu)\propto E_\nu^{-2}$ flux in Eq.~(\ref{eq:rate}) between $-3\%$ and $4\%$ with respect to the reference calculation of the exposure shown in Fig.~\ref{fig:exposure}. The uncertainty due to different models of $\nu_\tau$ cross-section and $\tau$ energy-loss affects mainly the ES channel with a corresponding range of variation of the event rate between $-28\%$ and $34\%$. The topography around the Observatory is not accounted for explicitly in the calculation of the exposure and is instead taken as a systematic uncertainty that would increase the event rate by an estimated $\sim\,20\,\%$ \cite{Auger_ES_PRL2008, Auger_ES_PRD2009}. The total uncertainty, obtained by adding these bands in quadrature, ranges from $-28\%$ to $39\%$, and is incorporated in the value of the limit itself through a semi-Bayesian extension \cite{Conrad} of the Feldman-Cousins approach \cite{Feldman-Cousins}.

The single-flavor $90\%$ C.L. integrated limit is:
\begin{equation}
k_{90}~< ~4.4 ~\times~ 10^{-9}~{\rm GeV~cm^{-2}~s^{-1}~sr^{-1}},
\label{eq:k90}
\end{equation}
or equivalently $1.4~{\rm EeV~km^{-2}~yr^{-1}~sr^{-1}}$. It mostly applies in the energy interval $10^{17}~{\rm eV} - 2.5\times10^{19}~{\rm eV}$ for which $\sim\,90\,\%$ of the total event rate is expected in the case of a $E_\nu^{-2}$ spectral flux. 
The relative contributions to the expected rate of events for a $E_\nu^{-2}$ flux due to the three neutrino flavors and to the ES and DG channels are displayed in Table~\ref{tab:relative_rates}. For such a spectral shape $\tau$ neutrinos contribute to $\sim\,86\,\%$ of the total event rate, and in particular ES neutrinos dominate the rate of $\nu_\tau$ events over the downward-going $\nu_\tau$. The contribution of $\nu_e$ and $\nu_\mu$ together is smaller than $15\,\%$ in this case.
\begin{table}[ht]
\begin{center}
\renewcommand{\arraystretch}{1.3}
\begin{tabular}{l  c} 
\hline
\hline
Flavor~           &  Relative contribution \\
\hline
$\nu_e$~          &   0.10      \\
$\nu_\mu$~        &   0.04      \\
$\nu_\tau$~       &   0.86      \\
\hline
\hline
Channel~                                   &   Relative contribution \\
\hline
Earth-skimming~$\nu_\tau$~                 &    0.79     \\
Downward-going~$\nu_e+\nu_\mu+\nu_\tau$~   &    0.21     \\
\hline
\hline
\end{tabular}
\end{center}

\caption{
Top of table: Relative contribution of the three neutrino flavors to the event rate in Auger due to a neutrino flux $\phi_\nu\propto E_\nu^{-2}$. 
Bottom: Relative contribution to the rate in the Earth-skimming (ES) and Downward-going (DG) channels. 
}
\label{tab:relative_rates}
\end{table}

\begin{figure*}[ht]
\centering
\includegraphics[width=0.9\textwidth]{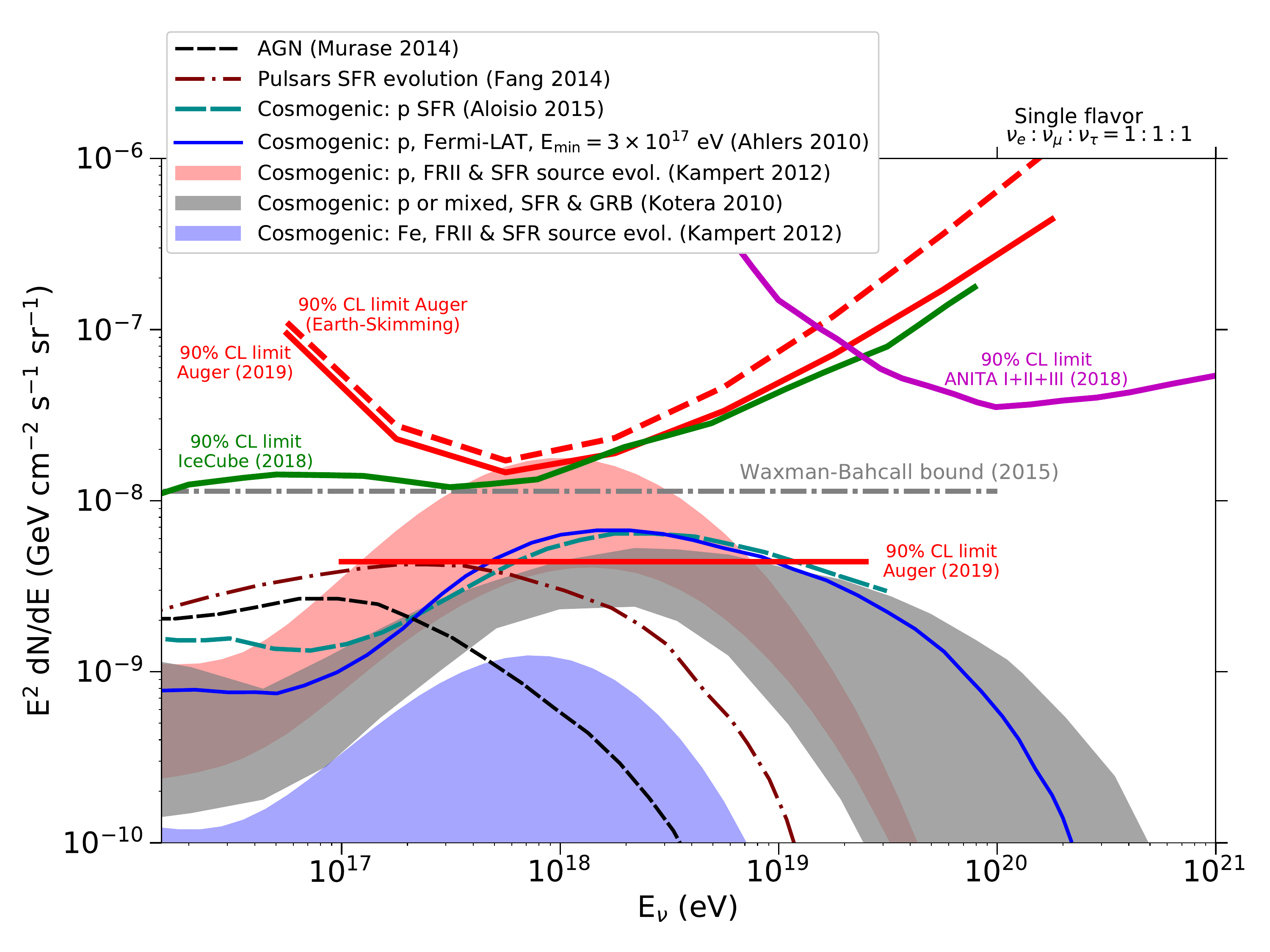}
\caption{ 
Pierre Auger Observatory upper limit (90$\%$ C.L.) to the normalization $k$ of the diffuse flux of UHE neutrinos $\phi_\nu = k~E_\nu^{-2}$ as given in Eqs.~(\ref{eq:k90a}) and (\ref{eq:k90}) (solid straight red line). Also plotted are the upper limits to the normalization of the diffuse flux (differential limits) when integrating the denominator of Eq.~(\ref{eq:k90a}) in bins of width 0.5 in $\log_{10}E_\nu$ (solid red line - Auger all channels and flavours; dashed red line - Auger Earth-skimming $\nu_\tau$ only). The differential limits obtained by IceCube \cite{IceCube_PRD2018} (solid green) and ANITA I+II+III \cite{ANITA_2018} (solid dark magenta) are also shown. 
The expected neutrino fluxes for several cosmogenic \cite{Kotera_GZK_2010, Ahlers_GZK_2010, Kampert_GZK_2012, Aloisio_GZK_2015} and astrophysical models of neutrino production, as well as the Waxman-Bahcall bound \cite{WB} are also plotted. All limits and fluxes are converted to single flavor.
}
\label{fig:differential}
\end{figure*}

The denominator of Eq.~(\ref{eq:k90a}) can also be integrated in bins of neutrino energy of width $\Delta E_\nu$, and a limit $\hat k_{90}$ can be obtained in each energy bin.
This is displayed in Fig.~\ref{fig:differential} for logarithmic energy intervals $\Delta\log_{10}E_\nu=0.5$. The differential limit is an effective way of characterizing the energy dependence of the sensitivity of a neutrino experiment. For the case of Auger it can be seen that the best sensitivity is achieved for energies around 1 EeV.

\section{Constraints on the origin of UHECR}
\label{sec:constraints}

With the upper limit obtained with the Observatory, we can constrain several classes of models of neutrino production in interactions of UHECR with the Cosmic-Microwave Background and Extragalactic-Background Light (EBL), often referred to as cosmogenic neutrino models.
The expected event rate in the Auger Observatory due to cosmogenic neutrinos depends strongly on the redshift evolution of the UHECR sources, on the nature of the primaries, namely whether they are protons or heavier nuclei, on the maximal redshift at which UHECR are accelerated, $z_{\rm max}$, and on the maximum energy acquired in the acceleration process, $E_{\rm max}$. Commonly, cosmogenic neutrino models assume that the observed UHECR flux-suppression \cite{HiRes_spectrum,Auger_spectrum} is based solely on the GZK effect, i.e.\ on energy losses of protons or nuclei in the CMB. The highest fluxes of cosmogenic neutrinos are then expected for injection of protons, while those expected for injection of iron nuclei are down typically by about an order of magnitude \cite{Hooper_GZK_2005,Ave_GZK_2005,Kotera_GZK_2010} (c.f.\ Fig. \ref{fig:differential}). We note, however, that the possibility of pure proton (or iron) primaries in the energy range of interest is disfavored by recent results on the composition of UHECR \cite{Auger_Xmax, Auger_Xmax_ICRC2017,Auger_Xmax_S1000, Auger_delta_SD}. Instead,  a gradually increasing fraction of heavier primaries is observed with increasing energy up to at least $E\sim 5\times 10^{19}$\,eV \cite{Auger_Xmax_ICRC2017}. In addition to this, adopting a simple astrophysical model fitting the energy spectrum and the mass composition suggests that the observed flux suppression is primarily an effect of the maximum rigidity of the sources of UHECR rather than only the effect of energy losses in the CMB and EBL \cite{Auger_combined-fit,Auger_composition_anisotropy}. In consequence, cosmogenic neutrino fluxes would be reduced much further and may escape detection for the foreseeable future \cite{Heinze19,Alves-Batista19,Wittkowski19}.
Thus, fluxes of cosmogenic neutrinos provide an independent probe of source properties and of the origin of the UHECR flux suppression at the highest energies.

In Table~\ref{tab:rates}, we show the expected number of events in the present lifetime of the Observatory for several cosmogenic neutrino models and the associated Poisson probability of observing no events. Scenarios assuming sources that accelerate only protons and that have a strong evolution with $z$, similar to Fanaroff-Riley type II (FRII) radio-loud AGN \cite{Kampert_GZK_2012}, are strongly constrained by the Auger results at more than $90\,\%$ C.L.
The increased exposure reported here allows us to constrain a larger parameter space of cosmogenic neutrino models than was possible previously \cite{Auger_nus_PRD2015}. 

If instead of protons, the primaries are heavier nuclei, photodisintegration is more likely than pion production, and the flux of cosmogenic neutrinos is suppressed \cite{Hooper_GZK_2005,Ave_GZK_2005,Kotera_GZK_2010}. For models assuming mixed (Galactic) composition of the UHECR such as those in \cite{Kotera_GZK_2010}, the constraints placed by the Observatory limits have improved with respect to previous publications \cite{Auger_nus_PRD2015} but they are still below the $90\,\%$ C.L.\ threshold of exclusion as can be seen in Table~\ref{tab:rates}. An about threefold increase in the current exposure will be needed to constrain at $90\,\%$ C.L.\ the lower edge of the gray band in  Fig.~\ref{fig:differential} for a mixed composition. The constraints are weaker if pure iron would be produced at the sources. Ruling out at $90\,\%$ C.L.\ the most optimistic predictions of cosmogenic neutrino flux at $10^{18}-10^{19}$\,eV in such a scenario \cite{Kampert_GZK_2012}, would require at least a sixfold increase in exposure. This is out of the range of sensitivity of the current configuration of Auger.

Another class of cosmogenic neutrino models \cite{Ahlers_GZK_2010} are constructed in such a way that the associated UHE photons produced along with the neutrinos, after cascading in the Universe and being converted into GeV photons, saturate measurements of the diffuse extragalactic $\gamma$-ray background  performed by \textit{Fermi}-LAT \cite{Fermi-LAT_diffuse}. 
Such approaches often neglect contributions from known populations of unresolved gamma-ray sources \cite{Ajello_ApJ2015,DiMauro_PRD2015}, and are also now in tension with the direct limits on cosmogenic neutrinos obtained with the Auger Observatory.

Also shown in Table~\ref{tab:rates} are some astrophysical models that are also excluded at $90\,\%$ C.L.\ such as those assuming UHECR acceleration to trans-GZK energies in radio-loud AGN \cite{Murase_AGN_2014}.

The IceCube Neutrino Observatory detected a flux of astrophysical neutrinos including a $\sim\,6.3$ PeV event compatible with being produced by the Glashow resonance \cite{Lu_UHECR2018, Halzen_ICRC2019}. Extrapolating the detected flux normalized at the energy of the Glashow event to Auger energies, using a simple power-law in energy $dN_\nu/dE_\nu \propto E^{-2.3}$, would yield $\sim\,0.4$ events in Auger, with an uncertainty in the normalization of the flux of a factor of $\sim\, 3$. This suggests that if the measured flux actually extends into the EeV energy range, it could be within reach of Auger in the next decade in an optimistic scenario.

\begin{table}[ht]
\begin{center}
\renewcommand{\arraystretch}{1.3}
\begin{tabular}{l c c} 
\hline
\hline
Neutrino Model   &  Expected number   & ~~~Probability of  \\
(Diffuse flux)   &  of $\nu$ events   & ~~~observing $0$   \\
\hline
\hline
Cosmogenic & & \\
\hline
\hline
(Kampert {\it et al.} \cite{Kampert_GZK_2012}) & & \\
proton, FRII      &  $\sim\,5.9$  & ~~~$\sim\,2.7\times 10^{-3}$ \\
proton, SFR       &  $\sim\,1.4$  & ~~~$\sim\,0.25$               \\
iron, FRII       &  $\sim\,0.4$  & ~~~$\sim\,0.67$ \\
\hline
(Aloisio {\it et al.} \cite{Aloisio_GZK_2015}) & & \\
proton, SFR      &  $\sim\,2.3$  & ~~~$\sim\,0.10$               \\
\hline
(Ahlers {\it et al.} \cite{Ahlers_GZK_2010}) & & \\
proton, $E_{\rm min}=10^{19}$ eV  &  ~~~$\sim\,4.6$  & $\sim\,1.0\times 10^{-2}$   \\
proton, $E_{\rm min}=10^{17.5}$ eV  &  ~~~$\sim\,2.4$  & $\sim\,9.0\times 10^{-2}$   \\
\hline
(Kotera {\it et al.} \cite{Kotera_GZK_2010}) & & \\
p or mixed, SFR \& GRB~             &  $\sim\,0.8~-~2.0$ & ~~~$\sim\,0.45~-~0.13$ \\
\hline
\hline
Astrophysical & & \\
\hline
\hline
(Murase {\it et al.} \cite{Murase_AGN_2014}) & & \\
Radio-loud AGN  &  $\sim\,2.9$ &  $\sim\,5.5 \times 10^{-2}$  \\
\hline
(Fang {\it et al.} \cite{Fang_Pulsars_2014}) & & \\
Pulsars - SFR    &  $\sim\,1.5$ & $\sim\,0.22$  \\
\hline
\hline
\end{tabular}
\end{center}
\caption{
Number of expected neutrino events $N_{\rm evt}$ in the period 1 Jan 2004 - 31 August 2018 for several models of UHE neutrino production (see Fig.~\ref{fig:differential}), given the exposure of the Surface Detector Array of the Pierre Auger Observatory shown in Fig.~\ref{fig:exposure}. The last column gives the Poisson probability $\exp({-N_{\rm evt}})$ of observing 0 events when the number of expected events is $N_{\rm evt}$.
}
\label{tab:rates}
\end{table}

\begin{figure*}[ht]
\centering
\begin{tabular}{c}
\includegraphics[width=0.88\textwidth]{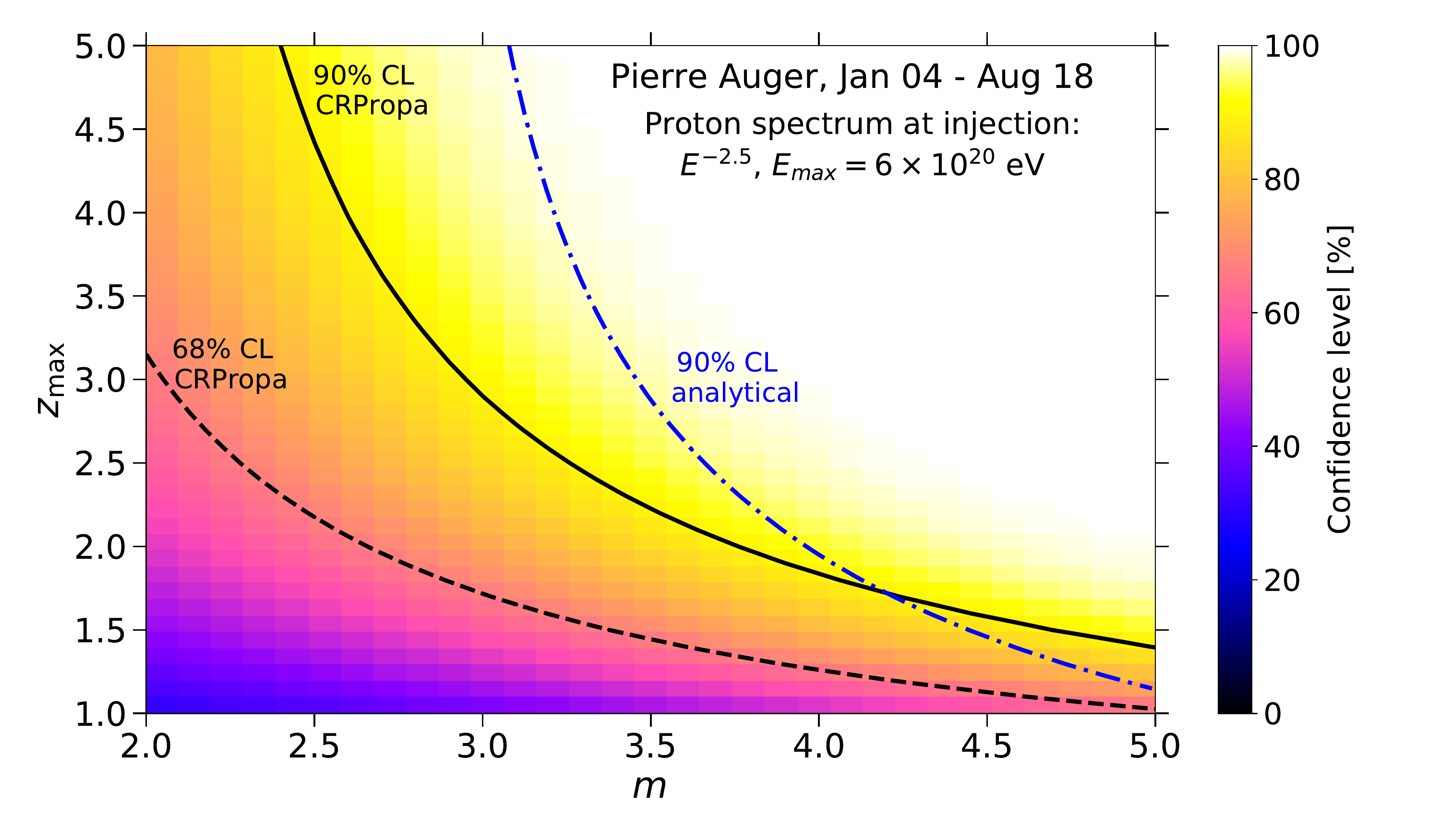} \\
\hspace{-2.1cm}\includegraphics[width=0.74\textwidth]{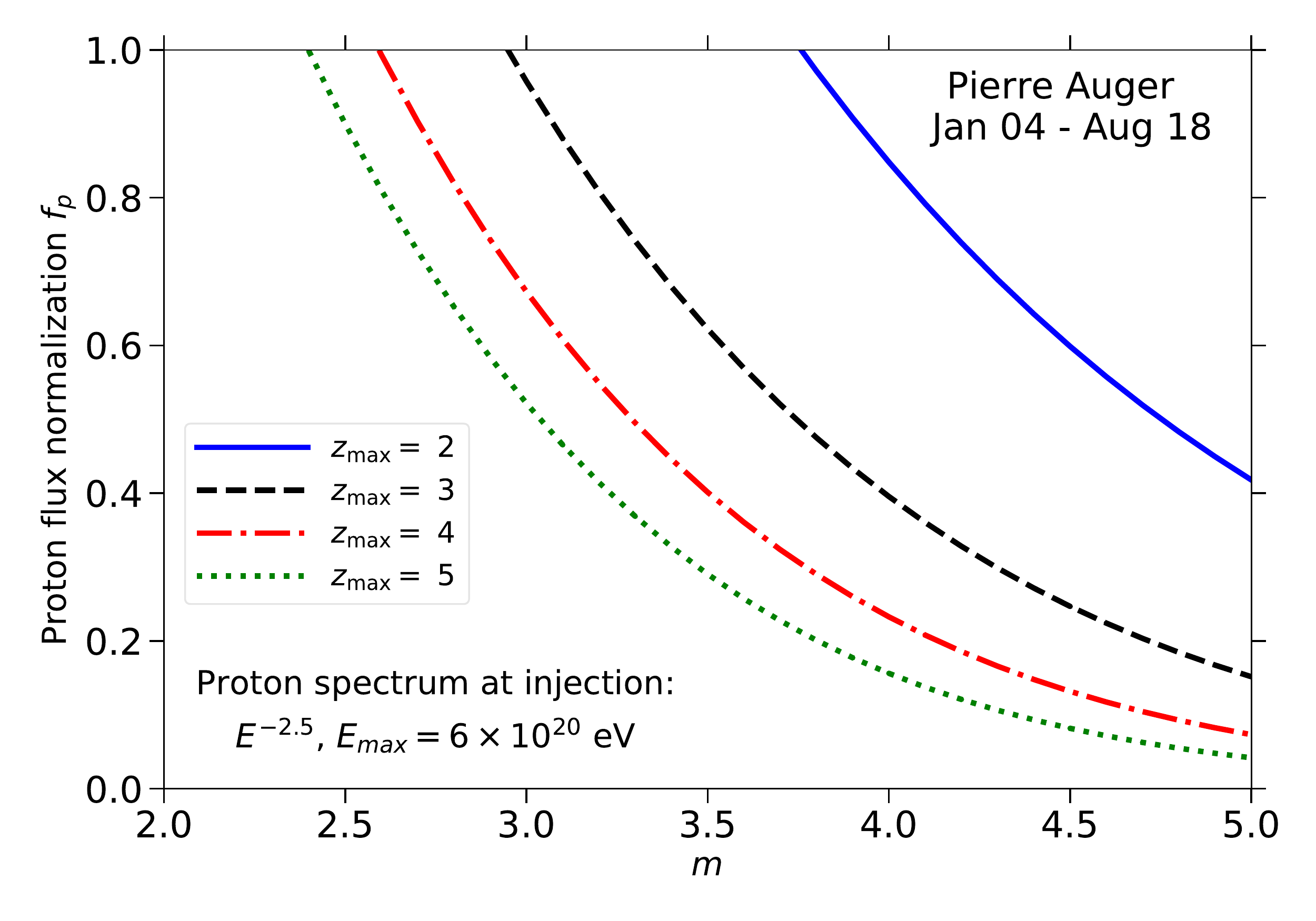} \\ 
\end{tabular}
\caption{ 
Constraints on UHECR source evolution models parameterized as $\psi(z)\propto (1+z)^m$ for sources distributed homogeneously up to a maximum redshift $z_{\rm max}$ and emitting protons following a power-law ${\rm d}N/{\rm d}E \propto E^{-2.5}$ up to $E=6 \times 10^{20}$\,eV. A proton-only flux is matched to the Auger spectrum at $7 \times 10^{18}$ eV (benchmark calculation for $f_p=1$, see text). The cosmogenic neutrino fluxes for each combination of $m$ and $z_{\rm max}$ were obtained with the Monte Carlo (MC) propagation code CRPropa \cite{CRPropa_2016}.
Top panel: Exclusion plot for source evolution parameter $m$ and $z_{\rm max}$ with $f_p=1$. The colored areas represent different levels of C.L.\ exclusion. In particular the solid and dashed lines represent the contours of $68\%$ and $90\%$ C.L.\ exclusion, respectively.
The dashed-dotted blue line represents the $90\%$ CL contour exclusion for cosmogenic neutrino models obtained with the analytical calculation in \cite{Yoshida_GZK_PRD2012}.
Bottom panel: Exclusion plot for source evolution model parameter $m$ and variable $f_p\leq 1$. The regions above the colored lines corresponding to several values of $z_{\rm max}$ are excluded at 90\,\% C.L.\ from the lack of neutrino candidates in Auger data.}

\label{fig:cosmogenic}
\end{figure*}

As demonstrated in Fig.~\ref{fig:differential}, the presently achieved upper bounds of cosmogenic neutrinos start to constrain astrophysical models that aim at describing the UHECR flux suppression above $4\times 10^{19}$ eV by energy losses of protons in the CMB. To further investigate the excluded parameter space of such proton models in more detail, we performed a comprehensive scanning of the evolution function of the sources with redshift, $\Psi(z)\propto (1+z)^m$ up to $z_{\rm max}$.
In all these scenarios, the spatial distribution of sources is assumed to be homogeneous, and all the sources are assumed to have the same UHECR luminosity. The simulations predicting the associated cosmogenic neutrino fluxes were performed with the CRPropa package \cite{CRPropa_2016} for a fixed spectral index of the UHECR proton spectrum $E^{-\alpha}$ at the sources (with $\alpha=2.5$), and a maximum energy of the protons $E_{\rm max} = 6\times 10^{20}$ eV. In the energy range $\sim \,3 \times 10^{17}$ to $\sim 3 \times 10^{18}$ eV, where the sensitivity of Auger peaks as shown in Fig.~\ref{fig:differential}, the cosmogenic neutrino flux is barely affected by the choice of $\alpha$ and $E_{\rm max}$ as long as the latter is not below $10^{20}$ eV (see Fig.\,1 in \cite{VanVliet_protons_GZKnus_2019}). These calculations include single- as well as multi-pion production in the proton-$\gamma$ interactions, and interactions with both the CMB and the EBL \cite{CRPropa_2016, VanVliet_protons_GZKnus_2019}. The proton flux at Earth is normalized to the Auger spectrum at $E=7 \times 10^{18}$ eV. For each pair of $m$ and $z_{\rm max}$, the cosmogenic flux was obtained, and the expected number of neutrino events in Auger was calculated. Those models predicting more than 2.39 neutrinos are disfavored at $> 90\,\%$ C.L. The resultant exclusion plot is shown in the top panel of Fig.~\ref{fig:cosmogenic} with the contours representing the $68\,\%$ and $90\,\%$ C.L.\ exclusion limits. The non-observation of neutrino candidates in the Observatory data allows us to disfavor a significant region of the parameter space $(m,z_{\rm max})$. This improves earlier constraints by the Pierre Auger Observatory \cite{Auger_nu_ICRC2017} based on a simplified analytical approach \cite{Yoshida_GZK_PRD2012}, also used by the IceCube neutrino observatory \cite{IceCube_PRL2016, IceCube_PRD2018}.

Finally, we explore the possibility of a proton component above $E \gtrsim 50$\,EeV \cite{Muzio_arXiv2019}, i.e.\ in the energy range presently not covered by the direct $X_{\mathrm{max}}$ measurements of Auger \cite{Auger_Xmax_ICRC2017}. Such a subdominant proton component would have a limited effect on the observed spectrum and composition, but would strongly alter the expected cosmogenic neutrino flux \cite{VanVliet_protons_GZKnus_2019}. This is because protons produce significantly more neutrinos when propagating through the universe than heavier nuclei of the same total energy, particularly if the latter were not accelerated significantly beyond the GZK energy threshold, such as indicated by a global fit to Auger data in \cite{Auger_combined-fit}. To be conservative, it is assumed that the cosmogenic neutrinos result only from the proton component. The rejection power achieved on such a model then depends on the relative contribution of protons normalized to the fixed all-particle flux denoted as $f_p$ 
\cite{VanVliet_protons_GZKnus_2019}. The corresponding exclusion plot is depicted in the bottom panel of Fig.\,\ref{fig:cosmogenic}. For instance, proton flux normalizations down to $f_p \simeq 0.2$ can be ruled out with the present data for sources following a cosmological evolution of $m\simeq 3.8$ up to redshift $z_{\rm max}=5$. 

The ongoing upgrade of the Surface Detector of the Pierre Auger Observatory, AugerPrime, will add a plastic scintillation detector \cite{AugerPrime} and a radio antenna \cite{RadioArray_ICRC2019} to each Water-Cherenkov Detector. Additional measurements of composition-sensitive observables will be possible with the data collected simultaneously with the Water-Cherenkov and the scintillation detectors for a high statistics sample of ultra-high energy events, not affected by the low duty cycle of the Fluorescence Detector. This will enhance the potential to identify the primary masses to the highest energies and will further add discrimination power for the primary mass on an event-by-event basis. It has been calculated that the upgrade will enable us to directly identify a possible proton component at the highest energies as small as $f_p=0.1$ \cite{AugerPrime,AugerPrime2}. This will be crucial for determining the role of cosmic-ray observations in UHE astronomy, and for establishing the potential of present and future detectors to the detection of the cosmogenic neutrino flux \cite{Alvarez-Muniz_ICRC17}.

\section*{Acknowledgments}

\begin{sloppypar}
The successful installation, commissioning, and operation of the Pierre
Auger Observatory would not have been possible without the strong
commitment and effort from the technical and administrative staff in
Malarg\"ue. We are very grateful to the following agencies and
organizations for financial support:
\end{sloppypar}

\begin{sloppypar}
Argentina -- Comisi\'on Nacional de Energ\'\i{}a At\'omica; Agencia Nacional de
Promoci\'on Cient\'\i{}fica y Tecnol\'ogica (ANPCyT); Consejo Nacional de
Investigaciones Cient\'\i{}ficas y T\'ecnicas (CONICET); Gobierno de la
Provincia de Mendoza; Municipalidad de Malarg\"ue; NDM Holdings and Valle
Las Le\~nas; in gratitude for their continuing cooperation over land
access; Australia -- the Australian Research Council; Brazil -- Conselho
Nacional de Desenvolvimento Cient\'\i{}fico e Tecnol\'ogico (CNPq);
Financiadora de Estudos e Projetos (FINEP); Funda\c{c}\~ao de Amparo \`a
Pesquisa do Estado de Rio de Janeiro (FAPERJ); S\~ao Paulo Research
Foundation (FAPESP) Grants No.~2010/07359-6 and No.~1999/05404-3;
Minist\'erio da Ci\^encia, Tecnologia, Inova\c{c}\~oes e Comunica\c{c}\~oes (MCTIC);
Czech Republic -- Grant No.~MSMT CR LTT18004, LO1305, LM2015038 and
CZ.02.1.01/0.0/0.0/16{\textunderscore}013/0001402; France -- Centre de Calcul
IN2P3/CNRS; Centre National de la Recherche Scientifique (CNRS); Conseil
R\'egional Ile-de-France; D\'epartement Physique Nucl\'eaire et Corpusculaire
(PNC-IN2P3/CNRS); D\'epartement Sciences de l'Univers (SDU-INSU/CNRS);
Institut Lagrange de Paris (ILP) Grant No.~LABEX ANR-10-LABX-63 within
the Investissements d'Avenir Programme Grant No.~ANR-11-IDEX-0004-02;
Germany -- Bundesministerium f\"ur Bildung und Forschung (BMBF); Deutsche
Forschungsgemeinschaft (DFG); Finanzministerium Baden-W\"urttemberg;
Helmholtz Alliance for Astroparticle Physics (HAP);
Helmholtz-Gemeinschaft Deutscher Forschungszentren (HGF); Ministerium
f\"ur Innovation, Wissenschaft und Forschung des Landes
Nordrhein-Westfalen; Ministerium f\"ur Wissenschaft, Forschung und Kunst
des Landes Baden-W\"urttemberg; Italy -- Istituto Nazionale di Fisica
Nucleare (INFN); Istituto Nazionale di Astrofisica (INAF); Ministero
dell'Istruzione, dell'Universit\'a e della Ricerca (MIUR); CETEMPS Center
of Excellence; Ministero degli Affari Esteri (MAE); M\'exico -- Consejo
Nacional de Ciencia y Tecnolog\'\i{}a (CONACYT) No.~167733; Universidad
Nacional Aut\'onoma de M\'exico (UNAM); PAPIIT DGAPA-UNAM; The Netherlands
-- Ministry of Education, Culture and Science; Netherlands Organisation
for Scientific Research (NWO); Dutch national e-infrastructure with the
support of SURF Cooperative; Poland -Ministry of Science and Higher
Education, grant No.~DIR/WK/2018/11; National Science Centre, Grants
No.~2013/08/M/ST9/00322, No.~2016/23/B/ST9/01635 and No.~HARMONIA
5--2013/10/M/ST9/00062, UMO-2016/22/M/ST9/00198; Portugal -- Portuguese
national funds and FEDER funds within Programa Operacional Factores de
Competitividade through Funda\c{c}\~ao para a Ci\^encia e a Tecnologia
(COMPETE); Romania -- Romanian Ministry of Research and Innovation
CNCS/CCCDI-UESFISCDI, projects
PN-III-P1-1.2-PCCDI-2017-0839/19PCCDI/2018 and PN18090102 within PNCDI
III; Slovenia -- Slovenian Research Agency, grants P1-0031, P1-0385,
I0-0033, N1-0111; Spain -- Ministerio de Econom\'\i{}a, Industria y
Competitividad (FPA2017-85114-P and FPA2017-85197-P), Xunta de Galicia
(ED431C 2017/07), Junta de Andaluc\'\i{}a (SOMM17/6104/UGR), Feder Funds,
RENATA Red Nacional Tem\'atica de Astropart\'\i{}culas (FPA2015-68783-REDT) and
Mar\'\i{}a de Maeztu Unit of Excellence (MDM-2016-0692); USA -- Department of
Energy, Contracts No.~DE-AC02-07CH11359, No.~DE-FR02-04ER41300,
No.~DE-FG02-99ER41107 and No.~DE-SC0011689; National Science Foundation,
Grant No.~0450696; The Grainger Foundation; Marie Curie-IRSES/EPLANET;
European Particle Physics Latin American Network; and UNESCO.
\end{sloppypar}


\newpage

{\bf\Large{The Pierre Auger Collaboration}}\\
\\

A.~Aab$^{75}$,
P.~Abreu$^{67}$,
M.~Aglietta$^{50,49}$,
I.F.M.~Albuquerque$^{19}$,
J.M.~Albury$^{12}$,
I.~Allekotte$^{1}$,
A.~Almela$^{8,11}$,
J.~Alvarez Castillo$^{63}$,
J.~Alvarez-Mu\~niz$^{74}$,
G.A.~Anastasi$^{42,43}$
% ...


\begin{description}[labelsep=0.2em,align=right,labelwidth=0.7em,labelindent=0em,leftmargin=2em,noitemsep]
\item[$^{1}$] Centro At\'omico Bariloche and Instituto Balseiro (CNEA-UNCuyo-CONICET), San Carlos de Bariloche, Argentina
\item[$^{2}$] Centro de Investigaciones en L\'aseres y Aplicaciones, CITEDEF and CONICET, Villa Martelli, Argentina
\item[$^{3}$] Departamento de F\'\i{}sica and Departamento de Ciencias de la Atm\'osfera y los Oc\'eanos, FCEyN, Universidad de Buenos Aires and CONICET, Buenos Aires, Argentina
\item[$^{4}$] IFLP, Universidad Nacional de La Plata and CONICET, La Plata, Argentina
\item[$^{5}$] Instituto de Astronom\'\i{}a y F\'\i{}sica del Espacio (IAFE, CONICET-UBA), Buenos Aires, Argentina
\item[$^{6}$] Instituto de F\'\i{}sica de Rosario (IFIR) -- CONICET/U.N.R.\ and Facultad de Ciencias Bioqu\'\i{}micas y Farmac\'euticas U.N.R., Rosario, Argentina
\item[$^{7}$] Instituto de Tecnolog\'\i{}as en Detecci\'on y Astropart\'\i{}culas (CNEA, CONICET, UNSAM), and Universidad Tecnol\'ogica Nacional -- Facultad Regional Mendoza (CONICET/CNEA), Mendoza, Argentina
\item[$^{8}$] Instituto de Tecnolog\'\i{}as en Detecci\'on y Astropart\'\i{}culas (CNEA, CONICET, UNSAM), Buenos Aires, Argentina
\item[$^{9}$] Observatorio Pierre Auger, Malarg\"ue, Argentina
\item[$^{10}$] Observatorio Pierre Auger and Comisi\'on Nacional de Energ\'\i{}a At\'omica, Malarg\"ue, Argentina
\item[$^{11}$] Universidad Tecnol\'ogica Nacional -- Facultad Regional Buenos Aires, Buenos Aires, Argentina
\item[$^{12}$] University of Adelaide, Adelaide, S.A., Australia
\item[$^{13}$] Universit\'e Libre de Bruxelles (ULB), Brussels, Belgium
\item[$^{14}$] Vrije Universiteit Brussels, Brussels, Belgium
\item[$^{15}$] Centro Brasileiro de Pesquisas Fisicas, Rio de Janeiro, RJ, Brazil
\item[$^{16}$] Centro Federal de Educa\c{c}\~ao Tecnol\'ogica Celso Suckow da Fonseca, Nova Friburgo, Brazil
\item[$^{17}$] Universidade de S\~ao Paulo, Escola de Engenharia de Lorena, Lorena, SP, Brazil
\item[$^{18}$] Universidade de S\~ao Paulo, Instituto de F\'\i{}sica de S\~ao Carlos, S\~ao Carlos, SP, Brazil
\item[$^{19}$] Universidade de S\~ao Paulo, Instituto de F\'\i{}sica, S\~ao Paulo, SP, Brazil
\item[$^{20}$] Universidade Estadual de Campinas, IFGW, Campinas, SP, Brazil
\item[$^{21}$] Universidade Estadual de Feira de Santana, Feira de Santana, Brazil
\item[$^{22}$] Universidade Federal do ABC, Santo Andr\'e, SP, Brazil
\item[$^{23}$] Universidade Federal do Paran\'a, Setor Palotina, Palotina, Brazil
\item[$^{24}$] Universidade Federal do Rio de Janeiro, Instituto de F\'\i{}sica, Rio de Janeiro, RJ, Brazil
\item[$^{25}$] Universidade Federal do Rio de Janeiro (UFRJ), Observat\'orio do Valongo, Rio de Janeiro, RJ, Brazil
\item[$^{26}$] Universidade Federal Fluminense, EEIMVR, Volta Redonda, RJ, Brazil
\item[$^{27}$] Universidad de Medell\'\i{}n, Medell\'\i{}n, Colombia
\item[$^{28}$] Universidad Industrial de Santander, Bucaramanga, Colombia
\item[$^{29}$] Charles University, Faculty of Mathematics and Physics, Institute of Particle and Nuclear Physics, Prague, Czech Republic
\item[$^{30}$] Institute of Physics of the Czech Academy of Sciences, Prague, Czech Republic
\item[$^{31}$] Palacky University, RCPTM, Olomouc, Czech Republic
\item[$^{32}$] Institut de Physique Nucl\'eaire d'Orsay (IPNO), Universit\'e Paris-Sud, Univ.\ Paris/Saclay, CNRS-IN2P3, Orsay, France
\item[$^{33}$] Laboratoire de Physique Nucl\'eaire et de Hautes Energies (LPNHE), Universit\'es Paris 6 et Paris 7, CNRS-IN2P3, Paris, France
\item[$^{34}$] Univ.\ Grenoble Alpes, CNRS, Grenoble Institute of Engineering Univ.\ Grenoble Alpes, LPSC-IN2P3, 38000 Grenoble, France, France
\item[$^{35}$] Bergische Universit\"at Wuppertal, Department of Physics, Wuppertal, Germany
\item[$^{36}$] Karlsruhe Institute of Technology, Institute for Experimental Particle Physics (ETP), Karlsruhe, Germany
\item[$^{37}$] Karlsruhe Institute of Technology, Institut f\"ur Kernphysik, Karlsruhe, Germany
\item[$^{38}$] Karlsruhe Institute of Technology, Institut f\"ur Prozessdatenverarbeitung und Elektronik, Karlsruhe, Germany
\item[$^{39}$] RWTH Aachen University, III.\ Physikalisches Institut A, Aachen, Germany
\item[$^{40}$] Universit\"at Hamburg, II.\ Institut f\"ur Theoretische Physik, Hamburg, Germany
\item[$^{41}$] Universit\"at Siegen, Fachbereich 7 Physik -- Experimentelle Teilchenphysik, Siegen, Germany
\item[$^{42}$] Gran Sasso Science Institute, L'Aquila, Italy
\item[$^{43}$] INFN Laboratori Nazionali del Gran Sasso, Assergi (L'Aquila), Italy
\item[$^{44}$] INFN, Sezione di Catania, Catania, Italy
\item[$^{45}$] INFN, Sezione di Lecce, Lecce, Italy
\item[$^{46}$] INFN, Sezione di Milano, Milano, Italy
\item[$^{47}$] INFN, Sezione di Napoli, Napoli, Italy
\item[$^{48}$] INFN, Sezione di Roma ``Tor Vergata'', Roma, Italy
\item[$^{49}$] INFN, Sezione di Torino, Torino, Italy
\item[$^{50}$] Osservatorio Astrofisico di Torino (INAF), Torino, Italy
\item[$^{51}$] Politecnico di Milano, Dipartimento di Scienze e Tecnologie Aerospaziali , Milano, Italy
\item[$^{52}$] Universit\`a del Salento, Dipartimento di Matematica e Fisica ``E.\ De Giorgi'', Lecce, Italy
\item[$^{53}$] Universit\`a dell'Aquila, Dipartimento di Scienze Fisiche e Chimiche, L'Aquila, Italy
\item[$^{54}$] Universit\`a di Catania, Dipartimento di Fisica e Astronomia, Catania, Italy
\item[$^{55}$] Universit\`a di Milano, Dipartimento di Fisica, Milano, Italy
\item[$^{56}$] Universit\`a di Napoli ``Federico II'', Dipartimento di Fisica ``Ettore Pancini'', Napoli, Italy
\item[$^{57}$] Universit\`a di Roma ``Tor Vergata'', Dipartimento di Fisica, Roma, Italy
\item[$^{58}$] Universit\`a Torino, Dipartimento di Fisica, Torino, Italy
\item[$^{59}$] Benem\'erita Universidad Aut\'onoma de Puebla, Puebla, M\'exico
\item[$^{60}$] Centro de Investigaci\'on y de Estudios Avanzados del IPN (CINVESTAV), M\'exico, D.F., M\'exico
\item[$^{61}$] Unidad Profesional Interdisciplinaria en Ingenier\'\i{}a y Tecnolog\'\i{}as Avanzadas del Instituto Polit\'ecnico Nacional (UPIITA-IPN), M\'exico, D.F., M\'exico
\item[$^{62}$] Universidad Aut\'onoma de Chiapas, Tuxtla Guti\'errez, Chiapas, M\'exico
\item[$^{63}$] Universidad Nacional Aut\'onoma de M\'exico, M\'exico, D.F., M\'exico
\item[$^{64}$] Institute of Nuclear Physics PAN, Krakow, Poland
\item[$^{65}$] University of \L{}\'od\'z, Faculty of Astrophysics, \L{}\'od\'z, Poland
\item[$^{66}$] University of \L{}\'od\'z, Faculty of High-Energy Astrophysics,\L{}\'od\'z, Poland
\item[$^{67}$] Laborat\'orio de Instrumenta\c{c}\~ao e F\'\i{}sica Experimental de Part\'\i{}culas -- LIP and Instituto Superior T\'ecnico -- IST, Universidade de Lisboa -- UL, Lisboa, Portugal
\item[$^{68}$] ``Horia Hulubei'' National Institute for Physics and Nuclear Engineering, Bucharest-Magurele, Romania
\item[$^{69}$] Institute of Space Science, Bucharest-Magurele, Romania
\item[$^{70}$] University Politehnica of Bucharest, Bucharest, Romania
\item[$^{71}$] Center for Astrophysics and Cosmology (CAC), University of Nova Gorica, Nova Gorica, Slovenia
\item[$^{72}$] Experimental Particle Physics Department, J.\ Stefan Institute, Ljubljana, Slovenia
\item[$^{73}$] Universidad de Granada and C.A.F.P.E., Granada, Spain
\item[$^{74}$] Instituto Galego de F\'\i{}sica de Altas Enerx\'\i{}as (I.G.F.A.E.), Universidad de Santiago de Compostela, Santiago de Compostela, Spain
\item[$^{75}$] IMAPP, Radboud University Nijmegen, Nijmegen, The Netherlands
\item[$^{76}$] KVI -- Center for Advanced Radiation Technology, University of Groningen, Groningen, The Netherlands
\item[$^{77}$] Nationaal Instituut voor Kernfysica en Hoge Energie Fysica (NIKHEF), Science Park, Amsterdam, The Netherlands
\item[$^{78}$] Stichting Astronomisch Onderzoek in Nederland (ASTRON), Dwingeloo, The Netherlands
\item[$^{79}$] Universiteit van Amsterdam, Faculty of Science, Amsterdam, The Netherlands
\item[$^{80}$] Case Western Reserve University, Cleveland, OH, USA
\item[$^{81}$] Colorado School of Mines, Golden, CO, USA
\item[$^{82}$] Department of Physics and Astronomy, Lehman College, City University of New York, Bronx, NY, USA
\item[$^{83}$] Louisiana State University, Baton Rouge, LA, USA
\item[$^{84}$] Michigan Technological University, Houghton, MI, USA
\item[$^{85}$] New York University, New York, NY, USA
\item[$^{86}$] Pennsylvania State University, University Park, PA, USA
\item[$^{87}$] University of Chicago, Enrico Fermi Institute, Chicago, IL, USA
\item[$^{88}$] University of Delaware, Department of Physics and Astronomy, Bartol Research Institute, Newark, DE, USA
\item[$^{89}$] University of Nebraska, Lincoln, NE, USA
\item[] -----
\item[$^{a}$] School of Physics and Astronomy, University of Leeds, Leeds, United Kingdom
\item[$^{b}$] Max-Planck-Institut f\"ur Radioastronomie, Bonn, Germany
\item[$^{c}$] Fermi National Accelerator Laboratory, USA
\item[$^{d}$] also at Universidade Federal de Alfenas, Po\c{c}os de Caldas, Brazil
\item[$^{e}$] Colorado State University, Fort Collins, CO, USA
\item[$^{f}$] now at Hakubi Center for Advanced Research and Graduate School of Science, Kyoto University, Kyoto, Japan
\item[$^{g}$] also at Karlsruhe Institute of Technology, Karlsruhe, Germany
\item[$^{h}$] also at Radboud Universtiy Nijmegen, Nijmegen, The Netherlands
\end{description}

\end{document}